\documentclass [a4paper,12pt]{article}

\usepackage{amsthm}
\usepackage{amsmath,amssymb,latexsym,amsfonts,mathrsfs}
\usepackage[dvipdfmx]{graphicx}

\usepackage{color}

\usepackage{mathtools}
\usepackage{fancyhdr}
\usepackage[top=30truemm, bottom=30truemm, left=25truemm, right=25truemm]{geometry}

\newcommand{\ep}{\varepsilon}
\newcommand{\nn}{\nonumber}

\newcommand{\SCR}[1]{{\mathscr #1}}

\newcommand{\CAL}[1]{{\cal #1}
}

\newcommand{\MAT}[1]{\left(\begin{array}{cccccccccc}#1\end{array}\right)}

\newcommand{\J}[1]{\left\langle #1 \right\rangle}
\newcommand{\D}[1]{{\mathscr D}( #1 )}

\newcommand{\CB}[1]{{ #1} }
\newcommand{\CG}[1]{{ #1} }

\theoremstyle{definition}
\newtheorem{Thm}{{\bf Theorem}}[section]

\newtheorem{Lem}[Thm]{{\bf Lemma}}
\newtheorem{Prop}[Thm]{{\bf Proposition}}

\newtheorem{Ass}[Thm]{{\bf Assumption}}

\newtheorem{Rem}[Thm]{{\bf Remark}}

\newcounter{Exami}

\newcommand{\Proof}[2][Proof]{
\begin{proof}[{\bf #1}]
#2
\end{proof}
}

\begin{document}

\begin{flushleft}
{\bf \Large Asymptotic behavior of solutions to nonlinear Schr\"{o}dinger equations with time-dependent harmonic potentials
} \\ \vspace{0.3cm}
by {\bf \large Masaki Kawamoto $^{\ast 1}$} and {\bf \large Ryo Muramatsu $^{\ast2}$}  \\
$^{\ast 1}$ Department of Engineering for Production, Graduate School of Science and Engineering, Ehime University, 3 Bunkyo-cho Matsuyama, Ehime, 790-8577. Japan \\ 
Email: kawamoto.masaki.zs@ehime-u.ac.jp \\
$^{\ast 2}$ Department of Mathematics, Graduate School of Science, Tokyo University
of Science, 1-3, Kagurazaka, Shinjuku-ku, Tokyo 162-8601, Japan. \\
Email: r.muramatsu728@gmail.com 
\end{flushleft}

\begin{center}
\begin{minipage}[c]{400pt}
{\bf Abstract}. {\small
In this study, we examine the asymptotic behavior of solutions to nonlinear Schr\"{o}dinger equations with time-dependent harmonic oscillators and prove the time-decay property of solutions in the case of a long range power type nonlinearity.
}
\end{minipage}
\end{center}

\begin{flushleft}
{\bf Keywords}; Nonlinear scattering theory; Long-range scattering; Time-dependent harmonic oscillators; Asymptotic behavior 
\end{flushleft}
\begin{flushleft}
\end{flushleft}
\section{Introduction}
{Throughout this paper}, we will consider nonlinear Schr\"{o}dinger equations with time-dependent harmonic potentials;
\begin{align}\label{eq1}
\begin{cases}
i \partial _t u(t,x) - \left( - \Delta /2 + \sigma (t) |x|^2/2 \right) u(t,x) = \nu F_{L}(u(t,x)) u(t,x) + \mu F_{S} (u(t,x)) u(t,x), \\
u(0,x) = u_0 (x),
\end{cases}
\end{align}
where $(t,x) \in {\bf R} \times {\bf R}^n$, $n \in \{ 1,2,3\}$, $ \nu , \mu \in {\bf R}$. $F_L:{{\bf C} \to {\bf R}}$ and $F_S :{{\bf C} \to {\bf R}}$ are nonlinear terms that are defined later. We assume the following assumption on the coefficient of harmonic oscillator $\sigma (t)$;
\begin{Ass} \label{A1}
Suppose $\sigma: {\bf R} \to {\bf R}$ and $ \sigma \in L^{\infty} ({\bf R})$, we define $y_1 (t)$ and $y_2(t)$ as linearly independent solutions to
\begin{align} \label{2}
y''_j(t) + \sigma(t) y_j(t) =0.
\end{align}
Then $y_1 (t)$ and $y_2 (t)$ satisfy the following conditions 
\begin{align*}
|y_2 (t) | \geq | y_1 (t) | \mbox{ as } |t| \gg 1, \quad \mbox{and }  \lim_{|t| \to \infty} \left| y_2 (t) \right| = \infty .
\end{align*}
Moreover, $y_1 (t)$, $y_2 (t)$, $y_1' (t)$, and $y_2' (t)$ are continuous functions.
\end{Ass}
\begin{Rem}
If $\sigma (t) = 0$, then we have $y_1 (t) = \mbox{const}$ and $y_2 (t) = t$ and if $\sigma (t) = -1$, then we have $y_1(t) = \CG{\sinh t}$ and $y_2(t) =\CG{ \cosh t}$. In the case where $\sigma (t)$ decays as $t^{2}\sigma(t) \to k$ with $0 \leq k <1/\CG{4}$, we obtain a linearly independent solution to \eqref{2} such that for $\lambda = (1-\sqrt{1-\CG{4k}})/2 \in [0,1/2) $ and for some constants $c_{1, \pm} \neq 0$ and $c_{2, \pm} \neq 0$, 
\begin{align*}
\lim_{ {t} \to \pm \infty} \frac{y_1(t)}{|t|^{ \lambda}}  =c_{1, \pm}, \quad 
\lim_{ {t} \to \pm \infty} \frac{y_2(t)}{|t|^{1- \lambda}}  =c_{2, \pm}
\end{align*}
hold. Models of $\sigma (t)$ for $k=0$ (i.e., $\lambda = 0$) can be observed in, e.g., Naito \cite{Na} and Willett \cite{W} and the models of $\sigma (t)$ for $\lambda \neq 0$ can be observed in, e.g., Geluk-Mari\'{c}-Tomi\'{c} \cite{GMT} (simplified models can be observed in Kawamoto \cite{Ka1} and Kawamoto-Yoneyama \cite{KY}).
\end{Rem}
In addition, we introduce some of the fundamental solutions to Hill's equation, which include
\begin{align*}
\zeta _j ''(t) + \sigma (t) \zeta _j (t) =0, \quad
\begin{cases}
\zeta _1 (0) = 1, \\
\zeta _1 ' (0) =0,
\end{cases}
\quad
\begin{cases}
\zeta _2 (0) = 0, \\
\zeta _2 ' (0) =1.
\end{cases}
\end{align*}

Here, let $v$ be a free solution associated with \eqref{eq1}, that is, $v$ is a solution to
\begin{align} \label{22}
\begin{cases}
i \partial _t v(t,x) + \left( \Delta /2 - \sigma (t)|x|^2/2 \right) v(t,x) = 0, \\
v(0,x) = v_0 \in L^2({\bf R}^n) \cap L^1({\bf R}^n), 
\end{cases}
\end{align}
and $U _0(t,s)$ be a propagator of $H_0(t) := - \Delta /2 + \sigma (t) |x| ^2/2$, that is, a family of unitary operators whose elements satisfy
\begin{align*}
i \frac{\partial}{\partial t} U_0(t,s) = H_0(t) U_0(t,s), \quad
i \frac{\partial}{\partial s} U_0(t,s) = - U_0(t,s) H_0(s), \quad U_0(s,s) = \mathrm{Id}_{L^2({\bf R}^n)}.
\end{align*}
Then, $v(t,\cdot) = U_0(t,0) v_0$ holds. Based on the results obtained in \cite{KY} (see Korotyaev \cite{Ko} and \cite{Ka3}), we derive the following dispersive estimate 
\begin{align}\label{disp}
\left\|
v(t, \cdot)
\right\| _{\infty} = \left\| U_0(t,0)v_0 \right\|_{\infty}  \leq C |\zeta _2 (t)|^{-n/2} \left\| v_0 \right\|_{1},
\end{align}
where $\| \cdot \|_{q}$, $1 \leq q \leq \infty$ denotes $\left\| \cdot \right\|_{L^{q}({\bf R}^n)}$. 
We now introduce the motivation for this paper. Let $ \phi (t,x)$ be a solution to \eqref{eq1} with $\sigma (t) \equiv 0$, $\mu =0$, and $F_L(u(t,x)) = |u(t,x)|^{\rho_0 -1}$ for some $\rho_0 >1$, and suppose $\phi (0,x)$ is included in some suitable function space. According to studies by Strauss \cite{S}, Barab \cite{Ba}, Tsutsumi-Yajima \cite{TY}, and so on, there exists $ \phi_{\pm} \in L^2({\bf R}^n) $ such that 
\begin{align} \label{3}
\lim_{t \to \pm \infty} \left\| \phi (t,\cdot) - e^{it\Delta /2} \phi_{\pm} \right\|_2 = 0
\end{align} 
holds for $1+2/n < \rho_0 $ and fails for $1< \rho_0 \leq 1+2/n$. Therefore, in this sense, when $\rho_0 > 1+2/n$ the nonlinearity is termed {\em short-range} and when $\rho_0 \leq 1 + {2/n}$ the nonlinearity is termed {\em long-range}. Therefore, the power $\rho_0 = 1+2/n$ denotes a threshold. For the case $\rho_0 = 1+2/n$, the long-range scattering theory has been considered in some papers, e.g., Hayashi-Ozawa \cite{HO}, Ozawa \cite{O}, Hayashi-Naumkin \cite{HN} among others. On the other hand, when we focus on the case where $t^2 \sigma (t) \to k \in [0,1/\CG{4})$ and $\zeta _2 (t) = \CAL{O}(|t|^{1- \lambda})$ for $|t| \gg 1$, it is expected that weak dispersive estimates 
\begin{align*}
\left\|
v(t, \cdot)
\right\| _{\infty} = \left\| U_0(t,0)v_0 \right\|_{\infty}  \leq C |\zeta _2 (t)|^{-n/2} \left\| v_0 \right\|_{1} \leq C |t|^{-n(1- \lambda)/2} \left\| v_0 \right\|_{1},
\end{align*}
will alter the threshold of nonlinearity from $1+2/n$ to $1 + 2/n(1-\lambda)$. Therefore, in this paper, we consider the manner in which $\sigma (t)$ affects the thresholds of nonlinearity. As an introduction to the main Theorems, we state the following conditions on the $\sigma (t)$; 
\begin{Ass}\label{A2}
For some $a_{1, \pm} \in {\bf R}$ and $a_{2, \pm} \neq 0$,
\begin{align*}
\lim_{t \to \pm \infty} \frac{\zeta_1(t)}{|y_2(t)|} = a_{1, \pm}, \quad
\lim_{t \to \pm \infty} \frac{\zeta_2(t)}{|y_2(t)|} = a_{2, \pm}
\end{align*}
hold. Moreover there exist $r_0 >0$ and $c>0$ such that for all $t \in (- \infty -r_0] \cup [r_0 , \infty )$, $| \zeta _2 (t) | >c$ holds.
\end{Ass}
If we consider the asymptotic behavior for the short-range case ($\nu =0$, $\mu \neq 0$) only, every arguments in \S{4} work well under the assumption \ref{A2}. However, when we consider long-range nonlinearities ($\nu \neq 0$), the decay condition $| \zeta _1 (t) / \zeta _2 (t) | \to 0$ as $|t| \to \infty$ acts very important role, this term appears in MDFM-decomposition \eqref{mdfm}. Indeed, by using such decay condition fully, the long-range scattering theory can be established (\cite{HO}, \cite{O}, \cite{HN} among others). To consider the case of $\nu \neq 0$ and to imitate approaches of previous works, we additionally assume the following conditions on $y_1$ and $y_2$.
\begin{Ass}\label{A3}
For some $b_{1, \pm} \neq 0$, $b_{2, \pm} \neq 0$ and $\delta _0 >0$ 
\begin{align*}
\lim_{t \to \pm \infty} \frac{\zeta_1(t)}{|y_1(t)|} = b_{1, \pm}, \quad
\lim_{t \to \pm \infty} \frac{\zeta_2(t)}{|y_2(t)|} = b_{2, \pm}, \quad \left| 
\frac{y_1(t)}{y_2(t)}
\right| \leq C |t|^{-\delta _0}
\end{align*}
holds. Moreover there exist $r_0 >0$ and $c>0$ such that for all $t \in (- \infty -r_0] \cup [r_0 , \infty )$, $| \zeta _2 (t) | >c$ holds.
\end{Ass}

 In this paper, we examine the asymptotic behavior of the solution to \eqref{eq1} using the approach of Hayashi-Naumkin \cite{HN}. To imitate this approach, we set for some $\gamma >0$, 
\begin{align*}
{H}^{\gamma ,0} := \left\{ u \in \SCR{S}'({\bf R}^n) \, | \, \left\|
u
\right\|_{\gamma ,0} := \left(
\int (1+|\xi|^2 )^{\gamma } \left| \hat{u} (\xi) \right| ^2 d \xi
\right) ^{1/2} < \infty\right\}
\end{align*}
and
\begin{align*}
{H}^{0, \gamma} := \left\{ u \in \SCR{S}'({\bf R}^n) \, | \, \left\|
u
\right\|_{0, \gamma} := \left(
\int (1+|x|^2)^{\gamma } \left| u (x) \right| ^2 d x
\right) ^{1/2} < \infty\right\}
\end{align*}
where $\hat{\cdot}$ indicates a Fourier transform. Finally, we present the following assumptions on nonlinearities; 
\begin{Ass}\label{A4}
Consider Assumption \ref{A1} and \ref{A2} or Assumption \ref{A1} and \ref{A3}. Let $\tilde{F}_L$, $\tilde{F}_S$: ${\bf R} \to {\bf R}$ satisfy the following 
\begin{align*}
\tilde{F}_S (\phi (t)) \leq C |\phi(t)|^{\rho_S}, \quad \tilde{F}_L (\phi (t)) \leq C |\phi (t)|^{\rho_L}, 
\end{align*}
where $\rho_S >0$ and $\rho_L >0$ satisfy 
\begin{align*}
\sup_{|t| \geq r_0} |\zeta _2 (t) |^{-n \rho_S/2} t^{1+ \delta _1} \leq C, \quad \sup_{|t| \geq r_0} |\zeta _2 (t) |^{-n \rho_L/2} t^{1} \leq C,
\end{align*}
for some $\delta _1 >0$. Assume that for some $  \gamma >n/2 $ and $\phi(t)=\phi(t,\cdot),\psi(t)=\psi(t,\cdot)$ there exists constant $C >0$ such that 
\begin{align} 
& F(\phi(t)) = F( |\phi (t)| )
\\ 
\label{gov2}
& \left\| 
F(\phi(t))
\right\|_{\infty} \leq C \tilde{F} ( \left\| \phi(t) \right\|_{\infty} ) ,
\\ 
 \label{gov} 
& \left\| 
F_{} (\phi(t)) \phi(t) 
\right\|_{\gamma ,0} \leq C  \tilde{F} (\left\|\phi(t) \right\|_{\infty} ) \left\| \phi(t) \right\|_{\gamma ,0}, 
\\ 
& \label{gov3}
\left\| 
 F_{} (\phi(t)) \phi(t) 
\right\|_{0, \gamma} \leq C \tilde{F} (\left\| \phi(t) \right\|_{\infty} ) \left\| \phi(t) \right\|_{0,\gamma}, 
\\ 
& \left\| F(\phi(t))\phi(t) -F(\psi(t))\psi(t)  \right\|_{k} \leq C \left( \tilde{F} (  \left\| \phi(t) \right\|_{\infty} )  + \tilde{F} ( \left\| \psi(t) \right\|_{\infty}) 
\right) \left\| {\phi(t)}-\psi(t) \right\|_{k} ,\\ 
\label{est}
& \left\| F(\phi(t))\phi(t) -F(\psi(t))\psi(t)  \right\|_{\gamma, 0} \leq C \left( \tilde{F} (  \left\| \phi(t) \right\|_{\infty} )  + \tilde{F} ( \left\| \psi(t) \right\|_{\infty}) 
\right) \left\| \phi(t)-\psi(t) \right\|_{\gamma, 0}  
\end{align} 
hold, where $k= \infty$ or $2$ and $(F, \tilde{F} ) = (F_L , \tilde{F}_L ) $ or $ (F_S , \tilde{F}_S) $.
\end{Ass}
We denote $F_S$ and $F_L$ as short-range nonlinearity and long-range nonlinearity, respectively. 
\begin{Rem}
We consider the case in which $\sigma (t)$ decays in $t$. The power type nonlinearities $F_S(u(t)) = |u(t)|^{2/n(1-\lambda) + \delta _2 } $ and $F_L(u(t)) = |u(t)|^{2/n(1-\lambda)}$ with some $\delta _2 >0$ that satisfies the assumption \ref{A4} (see, Lemma 2.1. and 2.3. in \cite{HN} ) can be considered examples. Because $\zeta _j(t) $ can be written as 
\begin{align} \label{8}
\zeta _j (t) = c_{j,1} y_1(t) + c_{j,2}y_2(t) 
\end{align} 
with constants $c_{j,1} , c_{j,2} \in{\bf R}$, $(c_{j,1},c_{j,2}) \neq (0,0)$, assumption \ref{A2} is the equivalent of assuming that $c_{2,2} \neq 0$ and assumption \ref{A1}. On the other hand, assumption \ref{A3} is the equivalent of assuming that $c_{1,2} = 0$, $c_{2,2} \neq 0$ and assumption \ref{A1}. In the case of short-range nonlinearity, it is enough to assume assumption \ref{A1} and \ref{A2} to obtain a dispersive estimate. On the other hand, the assumption \ref{A3} is needed to derive the dispersive estimates of long-range nonlinearity. The $\sigma (t)$, which satisfies assumption \ref{A1} and \ref{A2}, can be constructed with little difficulty; however, constructing the $\sigma (t)$ that satisfies \ref{A1} and \ref{A3} can be very difficult. When $\sigma (t)$ is non-continuous, we can construct $\sigma (t)$ as presented in \cite{KY}. We summarize several models that satisfy our assumptions; 
\begin{table}[htb]
  \begin{tabular}{l||c||r||r}
 $\sigma (t)$ & $ kt^{-2}$  & $t^2 \sigma (t) \to 0$ \mbox{ or } $0$ &  $-1$ \\ \hline 
 $\rho_S$ & $\rho_S > 2/(n(1-\lambda))$ & $\rho_S > 2/n$ & $\rho_S > \max ( n/2-1,0 ),  $ \\ \hline
 $\rho_L$ & $\rho_L = 2/(n(1- \lambda)) $ & ${\rho_L} = 2/n$ & $\times$ \\
  \end{tabular}
\end{table}
 \\ where  $ k\in [0,1/\CG{4})$ and $ \lambda = (1-\sqrt{1-\CG{4k}})/2$.
\end{Rem}
\begin{Rem}
If $f(u(t)) = F_S (u(t))$ or $F_L(u(t))$ satisfies $f(0) = f'(0) = 0$ and for all $z_1,z_2 \in {\bf C}$, 
\begin{align*}
\left| 
f'(z_1) -f'(z_2) 
\right|  \leq C 
\begin{cases}
|z_1-z_2| \cdot \CB{\max ( |z_1| ^{p-2},  |z_2|^{p-2} ) } &\mbox{if } p \geq 2, \\ 
|z_1-z_2| ^{p-1}  &\mbox{if } p \leq 2, 
\end{cases}
\end{align*} 
we obtain \eqref{gov} and \eqref{gov3} with $\tilde{f}(u(t)) = |u(t)|^{p-1} $, where $f' = \partial f/\partial z$ and $\partial f/ \partial \bar{z}$, $\tilde{f} = \tilde{F}_S$ or $\tilde{F}_L$ according to the result of Lemma 3.4. in Ginibre-Ozawa-Velo \cite{GOV}. Therefore, as for the power type nonlinearities such as $f(u(t)) = |u(t)|^{p-1}$ with some $p > \max ( 1, n/2 )$, the assumption \ref{A4} is then guaranteed with $n/2 < \gamma < p $. The log-like nonlinearity such as $f(u(t))= \tilde{f}(u(t)) = \left( \log (1+ 1/|u(t)|) \right)^{-(p-1)}$ with $p \geq 1+ {2/n}$ does not satisfy the assumptions presented in \ref{A4}, and therefore, we must relax the assumption \ref{A4} to include log-like nonlinearities such as nonlinearity suits when $\sigma (t) = -1$.
\end{Rem}

Consequently, we obtain the following dispersive estimates and asymptotics of the solution;
\begin{Thm}\label{thm}
Let $u_0 \in {H}^{\gamma,0} \cap {H}^{0, \gamma}$ with $ \gamma > n/2  $, $\left\| u_0 \right\|_{\gamma ,0} + \left\| u_0 \right\|_{0, \gamma} = \ep ' \leq \ep$, where $\ep >0$ is sufficiently small. If $\nu =0$, then under assumptions \ref{A1}, \ref{A2}, and \ref{A4}, there exists a unique global solution to \eqref{eq1} such that $u \in C({\bf R}; {H}^{\gamma ,0} \cap {H}^{0, \gamma}  )$ and 
\begin{align}\label{est1}
\left\|
u(t)
\right\|_{\infty} \leq C \ep ' (1+|\zeta _2 (t)|)^{-n /2}
\end{align}
hold, and if $\nu \neq 0$, then under assumptions \ref{A1}, \ref{A3}, and \ref{A4}, there exists a unique global solution to \eqref{eq1} such that $u \in C({\bf R}; {H}^{\gamma ,0} \cap {H}^{0, \gamma}  )$ and \eqref{est1} hold.
\end{Thm}
\begin{Thm}\label{T2}
Let $u_0 \in {H}^{\gamma,0} \cap {H}^{0, \gamma}$ with $ \gamma  > n/2 $, $\left\| u_0 \right\|_{\gamma ,0} + \left\| u_0 \right\|_{0, \gamma} = \ep ' \leq \ep$, where $\ep >0$ is sufficiently small, and $u \in C({\bf R}; H^{\gamma ,0} \cap H^{0, \gamma})$ is a global solution to \eqref{eq1}, which is presented in Theorem \ref{thm}. Moreover, consider the {assumptions \ref{A1}, \ref{A3}}, and \ref{A4}. Then there exist $W \in L^{\infty} \cap L^2$ and $\tilde{C}_1(\ep ' , \nu ) \geq 0$ and $ \tilde{C}_2 (\ep ' , \mu )  \geq 0$ with $ \tilde{C}_1(\ep ' , \nu ) = 0 $ iff $\nu=0$ and $ \tilde{C}_2 (\ep ' , \mu ) = 0$ iff $\mu =0$,  
\begin{align} \nn
& \left\| 
\SCR{F}\left( 
U_0(0,t) u(t, \cdot) \right) (t) \exp \left\{ 
i \nu \int_{r_0}^{t} F_L (|\zeta _2 (\tau) |^{-n/2} \SCR{F} (U_0 (0, \tau) u(\tau , \cdot )) ) d \tau
\right\} - W
\right\|_{k}  \\ & \leq  C \ep'  
t^{-\delta _0 \alpha + \tilde{C}_1(\ep ', \nu)  } + C \ep'  
t^{-\delta _1 + \tilde{C}_2(\ep ', \mu)  } \label{adrev1}
\end{align}
holds for $t \geq r_0$, where $k= 2$ or $\infty$, $\tilde{C}_1(\ep ', \nu) <\alpha < \min ( \gamma /2 -n/4, 1 )$ and $\delta _0$ and $\delta _1$ are equivalent to those in Assumption \ref{A3} and Assumption \ref{A4}, respectively.
\end{Thm}
\begin{Rem}
In the case where $t^{2} \sigma (t) \to 0$, one has $\delta _0 = 1$ and find that \eqref{adrev1} corresponds to the (1.4) in \cite{HN} with $n/2 < \gamma < 1 + 2/n$ and $F_L(u) = |u|^{2/n}$. Since the assumption $t^{2} \sigma (t) \to 0 $ includes $\sigma (t) \equiv 0$, our result may be a natural extension of result of \cite{HN}. On one hand, for $\sigma (t) = k t^{-2}$, $k \in [0,1/4 )$ with some additional assumptions (see, \cite{KY}), one has $| \zeta _2(t) / \zeta _1 (t) | / |t|^{1-2 \lambda} \to \mathrm{const} \neq 0$ as $|t| \to \infty$ with $\lambda = (1- \sqrt{1-4k})/2$. Then,
\begin{align*}
\left\|
u(t)
\right\|_{\infty} \leq C \ep ' (1 +|t|)^{-n(1- \lambda) /2}
\end{align*}
and 
\begin{align*}
&\left\| 
\SCR{F}\left( 
U_0(0,t) u(t, \cdot) \right) (t) \exp \left\{ 
i \nu \int_{r_0}^{t} F_L (|\zeta _2 (\tau) |^{-n/2} \SCR{F} (U_0 (0, \tau) u(\tau , \cdot )) ) d \tau
\right\} - W
\right\|_{k}  \\ & \leq  C \ep'  
t^{- (1-2 \lambda) \alpha + \tilde{C}_1(\ep ', \nu)  }  + C \ep'  
t^{-\delta _1 + \tilde{C}_2(\ep ', \mu)  }
\end{align*}
hold for $F_L (u) = |u|^{2/(n(1- \lambda))} $, $n/2 < \gamma < 1+ 2/(n(1- \lambda))$ and $\alpha < \min ( 1/2 + 1/(n(1- \lambda)) -n/4, 1 )$. On the other hand, for $\sigma (t) = -1$, $\zeta _1 (t) = \cosh t$ and $\zeta _2 (t) = \sinh t$ hold, and then letting $\nu = 0$, $F_S (u) = |u|^{\rho} $ with $\rho > \max (0, n/2 -1) $ and $n/2 < \gamma < 1 + \rho$, we find 
\begin{align*}
\left\|
u(t)
\right\|_{\infty} \leq C \ep ' (1 +|\sinh t|)^{-n/2}.
\end{align*}
\end{Rem}

In the case where $\sigma (t)$ decays in $t$, the second Theorem implies the threshold of short-range and long-range nonlinearities to be $2/n(1-\lambda)$. For $\sigma (t) = -1$, if we put $\nu =0$ and $F_S(u(t)) = |u(t)|^{\delta _2}$ with some $\delta _2 >0$, then we can obtain the Theorems with $\zeta _2(t) = \sinh (t)$. However, the short-range nonlinearity will ideally be $(\log (1+|u|^{-1}))^{-2/n - \delta _3}$ with some $\delta _3 >0$ because $F_S(|\zeta _2 (t)|^{-1}) \leq c |t|^{-n/2+ \delta _1}$ holds for such nonlinearity. To justify this argument, we must prove \eqref{gov} in the assumption \ref{A4} for this nonlinearity. 

In the above Theorems \ref{thm} and \ref{T2}, we reveal the asymptotic behavior of the solutions to \eqref{eq1} for both cases of long-range and short-range nonlinearities with generalized conditions of the coefficients of harmonic potential. Such results, particularly in the case where $\sigma(t)$ decays in $t$, are yet to be observed, and we are of the impression that this result is not mathematically and physically interesting. Similarly, Carles \cite{Ca} and Carles-Silva \cite{CaS} considered nonlinear Schr\"{o}dinger equations (NLS) with harmonic potential and time-dependent potentials, in these papers the global existence of solutions to \eqref{eq1} was proven under the more generalized potential including our models, and besides that they found the asymptotics of solutions of \eqref{eq1} for the case where $\sigma (t)$ satisfying $t^2 |\sigma (t)| < 1/4$ as $t \to \infty$. Moreover for the case where $\sigma (t) = -1$, \cite{Ca2} also considered the global well-posedness of the solution, scattering theory, and so on. On the other hand, the results of \cite{Ca} \cite{CaS} and \cite{Ca2} have not dealt with $L^{\infty}$ asymptotics, $L^2$ asymptotics for the case where $0 < \lim t^2 |\sigma (t)| \leq 1/4 $ and $H^{\gamma,0} \cap H^{0,\gamma}$-wellposedness with $\gamma > n/2$ ( $H^{1,0} \cap H^{0,1}$-wellposed can be found in \cite{CaS} even for the case $0 < \lim t^2 |\sigma (t)| \leq 1/4 $). We believe that this result is new and mathematically interesting. As a time-independent case, Hani-Thomann \cite{HT} proved the asymptotics in the case when $\sigma (t) \equiv \sigma \neq 0$. 

Our approach is applicable to different types of nonlinearities (see,e.g., Dodson \cite{D}, Masaki-Miyazaki-Uriya \cite{MMU}, Shimomura \cite{Si} and so on). Moreover, it has been determined that the approach used in \cite{HN} works well for the study on the lifespan of solutions to NLS (see, e.g., Sagawa-Sunagawa \cite{SS}), and hence our approach maybe applicable to such studies. 

In \S{3}, we found the $H^{\gamma,0} \cap H^{0,\gamma}$-wellposedness (local-in-time) of solution to \eqref{eq1}. The key approaches have been established by \cite{Ca} and \cite{CaS}, in particular the pseudo energy $\alpha (t)$ which acts very important roles were found by \S{5} of \cite{Ca}. In \S{2}, we summarize these and introduce some decomposition theorem of propagators. By combing their approaches and choosing suitable decompositions of propagators, we are finally able to find the  $H^{\gamma,0} \cap H^{0,\gamma}$-wellposedness, which is indispensable in order to imitate the approach of \cite{HN}. 

\section{Preliminaries and Auxiliary results}
In this section, we shall introduce some important lemma that appears in the proof . In this section, we always assume assumption \ref{A1}, \ref{A2}, and \ref{A4}. Throughout this paper, we use a constant $C>0$, which is always positive and is independent of any other parameters under consideration. We may use the notations 
\begin{align*}
 u(t, \cdot ) = u(t), \left( \mbox{resp. } v(t,\cdot) = v(t), \quad w(t,\cdot) = w(t) , \quad \mbox{and so on}  \right)
\end{align*} 
and 
\begin{align*}
G(u(t)) = \nu F_L(u(t)) u(t) + \mu F_S(u(t)) u(t).
\end{align*} 
We use the notation $p := -i \nabla$, $x^2 = |x| ^2$ and $p^2 = |p|^2 = - \Delta$. For $t$-depend functions $A_1(t)$ and $A_2 (t)$, the operator $| A_1 (t) p + A_2 (t) x| $ is defined as
\begin{align*}
\left| A_1 (t) p + A_2 (t) x \right| &= \left| e^{-iA_2(t) x^2/(2A_1(t))} \left( A_1(t) p \right) e^{iA_2(t) x^2/(2A_1(t))} \right| \\ &= e^{-iA_2(t) x^2/(2A_1(t))} \left| A_1(t) p \right| e^{iA_2(t) x^2/(2A_1(t))}.
\end{align*} 
To simplify the proof, we state the following lemma; 
\begin{Lem} \label{L01}
Let $a_1, a_2,a_3,a_4 \in{\bf R}$ and $\gamma >0$. Then for all $\phi \in L^2({\bf R}^n)$, there exists $C>0$ such that
\begin{align*}
\left\| \left( \left| a_1 p + a_2 x \right|^2 +a_3 p^2 + a_4 x^2 \right) ^{\gamma} (p^2+x^2+1)^{- \gamma} \phi \right\|_2  \leq C \left\| \phi \right\|_2
\end{align*}
and 
\begin{align*}
\left\| (p^2 + x^2+1)^{\gamma} (\J{p}^{ 2\gamma} + \J{x}^{2\gamma} )^{-1} \phi \right\|_2  \leq C 
\left\| \phi \right\|_2
\end{align*}
hold. 
\end{Lem}
This Lemma can be easily proved using the positiveness of harmonic oscillator $p^2 + x^2 \geq 1$ and Calder\'{o}n-Vaillancourt Theorem. By noting $p = -i \nabla _x $ and $x = i \nabla _p$, for constants $A_3$ and $A_4$,
\begin{align} \label{4}
\MAT{x \\ p} e^{-i A_3 x^2} = e^{-i A_3 x^2} \MAT{x \\ p-2A_3 x }, \quad
\MAT{x \\ p} e^{-i A_4 p^2} = e^{ -i A_4 p^2 }\MAT{x + 2A_4 p \\ p }.
\end{align}
Moreover, we denote $(x \cdot p + p\cdot x) /2$ by $A$. In the case of constant $A_5$, it  satisfies
\begin{align} \label{5}
\MAT{x \\ p} e^{-i A_5 A} = e^{-i A_5 A} \MAT{e^{A_5}x \\ e^{-A_5 } p }, 
\end{align}
 and the proof can be found in e.g., \S{2} of \cite{KY} and \S{2.3.} of \cite{Ka1}.
In addition, we occasionally use the notation $\J{\cdot} = (1+ \cdot ^2)^{1/2}$.

\subsection{Auxiliary results}

For the sake of using the approach presented in \cite{HN}, we must consider the Lemma 2.3 in \cite{HN} and the operator $J$ in the case when harmonic potential exists. 
To obtain these, we employ the formula in \cite{Ko};
\begin{Lem} The propagator $U_0(t,0)$ can be decomposed into the following form
\begin{align} \label{korotya}
U_0(t,0) = e^{i \zeta _1 '(t) x^2/(2 \zeta _1 (t))} e^{-i (\log |\zeta _1 (t) | ) A}e^{-i \zeta _2 (t) p^2 /(2 \zeta _1(t))} {\bf S}^{\nu (t)} ,
\end{align}
where $({\bf S} f) (x) = e^{-in\pi/2} f(-x)$ for $f\in L^2({\bf R}^n)$ and $\nu(t)$ is the number of zeros in the elements of
$\left\{ \tau \in [0,t] \, | \, \zeta _1(\tau) =0 \right\}$ for $t \geq 0$ or $\left\{ \tau \in [t,0] \, | \, \zeta _1(\tau) =0 \right\}$ for $t \leq 0$,
(see \cite{Ko} and \cite{KY}).
\end{Lem}

Based on the operator calculation (see, e.g., below (2.2) of \cite{KY}), we obtain
\begin{align*}
\left( e^{-i (\log |\zeta _1 (t)|) A } v \right) (x) = |\zeta _1 (t)|^{-n/2} v(|\zeta _1 (t)|^{-1} x ).
\end{align*}
Using this lemma, the following proposition holds;
\begin{Prop}\label{prop2.2}
Define $f(u(t))$ and $\tilde{f}(u(t))$ as either $F_S(u(t))$ and $\tilde{F}_S (u(t))$ or $F_L(u(t))$ and $\tilde{F}_L (u(t))$, respectively. Let $\gamma  > 0$, then
\begin{align}\label{est22}
\left\| U_0(0,t) f(u(t)) u(t) \right\|_{{0, \gamma}} \leq C \tilde{f} (  \left\| u(t) \right\|_{{\infty} } )\left\| U_0(0,t) u(t) \right\| _{{0, \gamma}}
\end{align}
holds for all $t\in \{t \in {\bf R} \, | \, \zeta _1 (t) \neq 0 \} $.
\end{Prop}
\Proof{
By \eqref{korotya}, we have
\begin{align*}
U_0(0,t) &= \left( {\bf S}^{\nu (t)}\right)^{-1}e^{i \zeta _2 (t) p^2 /(2 \zeta _1(t))}e^{i (\log |\zeta _1 (t) | ) A}e^{ -i \zeta _1 '(t) x^2/(2 \zeta _1 (t))} \\ &=
e^{i \zeta _2 (t) p^2 /(2 \zeta _1(t))}e^{i (\log |\zeta _1 (t) | ) A}e^{ -i \zeta _1 '(t) x^2/(2 \zeta _1 (t))} \left( {\bf S}^{\nu (t)}\right)^{-1}, 
\end{align*}
where ${\bf S}$ commutes with $p^2$, $A$, and $x^2$. We define $v(t,x) = e^{ -i \zeta _1 '(t) x^2/(2 \zeta _1 (t))} ({\bf S} ^{\nu (t)})^{-1}u(t,x) $ and $w(t,x) := v(t,|\zeta _1(t)| x )$. Then using lemma 2.3. of \cite{HN} and assumption \ref{A4}, 
\begin{align*}
\left\| |x|^{\gamma}
U_0(0,t) f(u(t)) u(t)
\right\|_{2 } &= |\zeta _1 (t)|^{n/2} \left\|
|x|^{\gamma} e^{i \zeta _2 (t) p^2 /(2 \zeta _1(t))} f(w(t)) w(t)
\right\|_2 \\ & 
\leq C |\zeta _1 (t)|^{n/2}\left\| 
 {f}( w(t) ) \right\|_{\infty}  \left\| |x|^{\gamma} e^{i \zeta _2 (t) p^2 /(2 \zeta _1(t))}  w(t) \right\|_2
\\ & \leq C
|\zeta _1 (t)|^{n/2} \tilde{f} ( \left\| w(t) \right\|_{\infty}^{\rho -1} ) \left\|
|x|^{\gamma} e^{i \zeta _2 (t) p^2 /(2 \zeta _1(t))} w(t)
\right\|_2.
\end{align*}
By using $|\zeta _1 (t)|^{n/2} w(t) = e^{i (\log |\zeta _1 (t) | ) A}e^{ -i \zeta _1 '(t) x^2/(2 \zeta _1 (t))} \left( {\bf S}^{\nu (t)}\right)^{-1} u(t) $ and $\| w (t) \|_{\infty} = \| u (t) \|_{\infty}$, we have \eqref{est22}.
}

\subsection{Pseudo energy}
Using proposition \ref{prop2.2}, we obtain the estimate for $\left\| \cdot \right\|_{0,\gamma}$, which played a very important role in \cite{HN}. However, it is proven only on the region $t\in \{t \in {\bf R} \, | \, \zeta _1 (t) \neq 0 \} $ and hence we now prove the same statement of Proposition \ref{prop2.2} on ${\bf R} \backslash \{t \in {\bf R} \, | \, \zeta _1 (t) \neq 0 \}$. Now we introduce $\alpha (t)$ here and notice that we can observe that this $\alpha (t)$ exhibits this good relation, see also \S{5} of \cite{Ca};
\begin{Lem}
Define the operator $\alpha (t)$ as
\begin{align*}
\alpha (t) := \left( \zeta _2 (t) p - \zeta _2 '(t) x \right)^2.
\end{align*}
Then for all $(t,s) \in {\bf R}^2$ and $\gamma >0$,
\begin{align*}
\J{\alpha (t)}^{\gamma} U_0(t,s) = U_0(t,s) \J{\alpha (s)}^{\gamma}
\end{align*}
holds, where $\J{\cdot} = (1+ \cdot ^2)^{1/2}$. In particular, by remarking $\zeta _2'(0) =1 $ and $\zeta _2 (0) =0$,
\begin{align*}
\J{x}^{\gamma} U_0(0,t) =  \J{
\zeta _2 (0) p - \zeta _2 '(0) x}^{\gamma} U_0(0,t) = U_0(0,t) \J{\alpha (t)}^{\gamma} .
\end{align*}
\end{Lem}
\Proof{
We first state that ${\bf S}$ commutes with $p^2$, $A$ , $x^2$, and $\alpha (t) $. We then have
\begin{align*}
\alpha (t) U_0(t,s) = ({\bf S}^{\nu (t)}) ( {\bf S}^{\nu (s)})^{-1} \alpha (t) \tilde{U}_0 (t,0) \tilde{U}_0(0,s),
\end{align*}
where $\tilde{U}_0(t,s) =({\bf S}^{\nu (s)})  ({\bf S}^{\nu (t)} )^{-1} U_0(t,s) $. Now we prove $\alpha (t) \tilde{U}_0(t,0) = \tilde{U}_0 (t,0) x^2 $. Indeed, by noting \eqref{4}, \eqref{5}, and that 
\begin{align*}
\zeta _1 (t) \zeta _2 '(t) - \zeta _1 '(t) \zeta _2 (t) = 1, 
\end{align*}
we have 
\begin{align*}
\alpha (t) \tilde{U}_0(t,0) &= e^{i \zeta _1 '(t) x^2/(2 \zeta _1 (t))}\left(
\zeta _2 (t) p - \frac{1}{\zeta _1 (t)} x
\right) ^2 e^{-i A \log |\zeta _1 (t)| } e^{-i \zeta _2(t) p^2/(2 \zeta _1 (t))} \\ &=
e^{i \zeta _1 '(t) x^2/(2 \zeta _1 (t))}e^{-i A \log |\zeta _1 (t)| }
\left(
\frac{\zeta _2 (t)}{\zeta _1 (t)} p - x
\right) ^2 e^{-i \zeta _2(t) p^2/(2 \zeta _1 (t))} \\ &=
\tilde{U}_0 (t,0) (-x) ^2.
\end{align*}
Next, we prove that $x^2 \tilde{U}_0(0,s) = \tilde{U}_0 (0,s) \alpha (s)$. Indeed,
\begin{align*}
x^2 \tilde{U}_0 (0,s) &= e^{i \zeta _2 (s) p^2/(2 \zeta _1 (s))} \left(
x- \frac{\zeta _2 (s)}{\zeta _1 (s)} p
\right) ^2 e^{i \log |\zeta _1 (s) | A} e^{-i \zeta _1 '(s)x^2/(2 \zeta _1 (s)) } \\ &=
e^{i \zeta _2 (s) p^2/(2 \zeta _1 (s))} e^{i \log |\zeta _1 (s) | A}
\left(
\frac{x}{\zeta _1 (s)} - \zeta _2 (s) p
\right) ^2 e^{-i \zeta _1 '(s)x^2/(2 \zeta _1 (s)) } \\ &=
\tilde{U}_0(0,s) \alpha (s), 
\end{align*}
which proves 
\begin{align*}
\alpha (t) \tilde{U}_0(t,s) = \alpha (t) \tilde{U}_0(t,0) \tilde{U}_0(0,s) = \tilde{U}_0(t,s) \alpha (s)
\end{align*}
By using 
\begin{align*}
\J{\alpha (t)}^{\gamma} \tilde{U}_0(t,s) = \left( 1 +  \tilde{U}_0(t,s) \alpha (s)^2  \tilde{U}_0(s,t) \right)^{\gamma /2} \tilde{U}_0(t,s) = \tilde{U}_0(t,s) \J{\alpha (s)}^{\gamma}, 
\end{align*}
we have Lemma.
}
\begin{Rem}
For to simplify the proof, we use the Korotyaev's decomposition formula in the proof of this lemma. However, this lemma can be proven without using Korotyaev's decomposition formula but with using commutator calculation. 
\end{Rem} 
In this paper, we present $\alpha (t)$ pseudo-energy in $t$. We then obtain the following important estimate; 
\begin{Prop}\label{P1}
We define $f(u(t))$ and $\tilde{f}(u(t))$ as either $F_S(u(t))$ and $\tilde{F}_S (u(t))$ or $F_L(u(t))$ and $\tilde{F}_L (u(t))$, respectively. Let $ \gamma >0 $, then for all $t \in \{ t \in {\bf R} \, | \, \zeta _2 (t) \neq 0 \} $, 
\begin{align}
 \left\| \J{\alpha (t)}^{\gamma /2} f(u(t)) u(t) \right\|_{2}  \leq C \tilde{f} ( \left\| u(t) \right\|_{{\infty} }) \left\| \J{\alpha (t)}^{\gamma /2} u(t) \right\| _{2} \label{est2}
\end{align} 
and 
\begin{align}
\left\| U_0(0,t) f(u(t)) u(t) \right\|_{0,\gamma} \leq C \tilde{f} ( \left\| u(t) \right\|_{{\infty} }) \left\| U_0(0,t)u(t) \right\| _{0,\gamma} \label{est2'}
\end{align}
hold.
\end{Prop}
\Proof{
Because $\J{x}^2 {U_0(0,t)} = {U_0(0,t)} \J{\alpha (t)}$ and $U_0(0,t)$ is the unitary operator on $L^2({\bf R}^n)$, \eqref{est2'} is proven as a sub-consequent of \eqref{est2}, and therefore, we only prove \eqref{est2}. Using 
\begin{align*}
|\alpha (t)|^{\gamma /2} = |\zeta _2 (t)|^{\gamma}e^{i \zeta _2 '(t)x^2/(2\zeta _2(t)} | p |^{\gamma} e^{-i \zeta _2 '(t)x^2/(2\zeta _2(t))}
\end{align*}
and 
\begin{align*}
e^{-i \zeta _2 '(t)x^2/(2\zeta _2(t))} f(u (t)) u(t) =  f\left(  e^{-i \zeta _2 '(t)x^2/(2\zeta _2(t))} u(t) \right) e^{-i \zeta _2 '(t)x^2/(2\zeta _2(t))} u(t), 
\end{align*}
for $l(t) =e^{-i \zeta _2 '(t)x^2/(2\zeta _2(t))} u(t) $ we have 
\begin{align*}
\left\| |\alpha (t)|^{\gamma /2} f(u(t)) u(t) \right\|_{2} = |\zeta _2 (t)|^{\gamma} \left\| |p|^{\gamma} f(l(t)) l(t) \right\|_{{2}}. 
\end{align*} 
Then by assumption \ref{A4}, we have 
\begin{align*}
\left\| |\alpha (t)|^{\gamma /2} f(u(t)) u(t) \right\|_{2}  &\leq C |\zeta _2(t)|^{\gamma} \tilde{f} \left( \left\| l(t)\right\| _{\infty} \right) \left\| |p|^{\gamma} l(t)\right\|_2 \\&= 
C \tilde{f} \left( \left\| u(t) \right\|_{{\infty} } \right) \left\| |\alpha (t)|^{\gamma /2} u(t) \right\| _{2}.
\end{align*} 
}
\begin{Prop}\label{P01}
We define $f(u(t))$ and $\tilde{f}(u(t))$ as either $F_S(u(t))$ and $\tilde{F}_S (u(t))$ or $F_L(u(t))$ and $\tilde{F}_L (u(t))$, respectively. Let $ \gamma >0 $. Then for all $t \in {\bf R} $, 
\begin{align}
\left\| U_0(0,t) f(u(t)) u(t) \right\|_{0,\gamma} \leq C \tilde{f} ( \left\| u(t) \right\|_{{\infty} }) \left\| U_0(0,t)u(t) \right\| _{0,\gamma} \label{est2'}
\end{align}
holds.
\end{Prop}
\Proof{
It is enough to prove that $\{ t \in {\bf R} \, | \, \zeta _1 (t) = 0 \} \cap \{ t \in {\bf R} \, | \, \zeta _2 (t) = 0 \} = \emptyset $ but this clearly holds from $\zeta _1 (t) \zeta _2 '(t) - \zeta _1'(t) \zeta _2 (t) =1$ for all $t \in {\bf R}$ holds.
}

\subsection{MDFM decomposition.}
In this subsection, we introduce the so called MDFM decomposition of $U_0(t,0)$. A non-singular type decomposition (MDMDFM decomposition) was presented in \cite{Ko}, and this decomposition was modified by Adachi-Kawamoto \cite{AK}, and Kawamoto \cite{Ka3}. However, to imitate the approach of \cite{HN}, we must obtain the MDFM decomposition. To consider this issue, the following lemma, which was obtained by Kawamoto \cite{Ka2}, is very useful;
\begin{Lem}
Let $a(t)$, $b(t)$ and $c(t)$ be
\begin{align*}
a(t) = \frac{1-\zeta _2 '(t)}{2 \zeta _2 (t)}, \quad b(t) = \frac{\zeta _2 (t)}{2} ,\quad
c(t) = \frac{1-\zeta _1 (t)}{2 \zeta _2 (t)}.
\end{align*}
Then the following decomposition of the propagator $U_0(t,0)$ holds;
\begin{align}\label{decomp}
U_0(t,0) = e^{-ia(t) x^2} e^{-ib(t) p^2} e^{-ic(t) x^2}.
\end{align}
\end{Lem}
\begin{Rem}
In \cite{Ka2}, only the case where $\sigma (t) \geq 0$ was considered. However, every argument that proves \eqref{decomp} works even if $\sigma (t)$ is negative.
\end{Rem} 
Using this lemma, we obtain the MDFM-decomposition;
\begin{Thm}
For $\phi \in \SCR{S}({\bf R}^n)$, let us define
\begin{align*}
\left( \CAL{M}(\tau) \phi \right) (x) = e^{ix^2/(2 \tau)} \phi (x), \quad
\left(
\CAL{D}(\tau) \phi
\right) (x) = \frac{1}{(i \tau)^{n/2}} \phi (x/ \tau).
\end{align*}
Then the following MDFM decomposition holds;
\begin{align}\label{mdfm}
U_0(t,0) = \CAL{M} \left(  \frac{\zeta _2(t)}{\zeta _2 '(t)} \right) \CAL{D} (\zeta _2 (t)) \SCR{F} \CAL{M} \left( \frac{\zeta _2 (t)}{ \zeta _1 (t)} \right)
\end{align}
\end{Thm}
\Proof{
For $\phi \in \SCR{S}({\bf R}^n)$,
\begin{align*}
(U_0(t,0) \phi )(x) &= e^{-ia(t) x^2}\frac{1}{(4 \pi i b(t) )^{n/2}} \int
e^{i(x-y)^2/(4b(t))} e^{-ic(t) y^2} \phi (y) dy
\\ &= e^{-i(a(t)-1/(4b(t)) ) x^2} \CAL{D}(2b(t)) \SCR{F}[\psi] (x),
\end{align*}
where $\psi (y) = e^{i(1/(4b(t))-c(t) )y^2 }\phi (y) $. Together with
\begin{align*}
a(t) - \frac{1}{4b(t)} = -\frac{\zeta _2 '(t)}{2 \zeta _2 (t)}, \quad
\frac{1}{4b(t)} -c(t) = \frac{\zeta _1(t)}{ 2 \zeta _2 (t)},
\end{align*}
we can obtain \eqref{mdfm}.
}

Using this lemma, we can prove the $\| \cdot \|_{\infty}$ decay estimate;
\begin{Lem}\label{lem2.6}
Let $u(t,x)$ be a smooth function and $|t| \geq r_0$. Then under assumption \ref{A1} and \ref{A2},
\begin{align}\label{est3}
\left\|
u(t)
\right\|_{\infty} \leq C |\zeta _2 (t)|^{-n/2} \left\|
\SCR{F} U_0(0,t) u(t)
\right\|_{\infty} + C |\zeta _2 (t)|^{-n/2} \left| \frac{\zeta _1 (t)}{\zeta _2 (t)} \right|^{ \alpha} \left\|
U_0(0,t) u(t)
\right\|_{0, \gamma}
\end{align}
holds for $|t| \geq r_0$, where $\alpha \in (0,1)$ and $\gamma > n/2 + 2 \alpha$. 
\end{Lem}
\Proof{
The proof can be obtained by imitating the approach used in Lemma 2.2. of \cite{HN}. The identity $u(t) = U_0(t,t) u(t)$ and \eqref{decomp} yields
\begin{align*}
u(t) &= \frac{e^{-ia(t) x^2}}{(4\pi i b(t))^{n/2}} \int e^{i(x-y)^2/(4b(t))} e^{-ic(t) y^2} (U_0(0,t) u)(y) dy \\ &=
\frac{e^{i (\zeta _2 '(t)/(2 \zeta _2 (t)) ) x^2}}{(4\pi i b(t))^{n/2}} \int e^{-i x \cdot y /(2b(t))} \left\{
e^{i (\zeta _1(t)/(2 \zeta _2 (t)) ) y^2} -1 +1
\right\}(U_0(0,t) u)(y) dy \\ &=
\frac{e^{i (\zeta _2 '(t)/(2 \zeta _2 (t)) ) x^2}}{(2 i b(t))^{n/2}} \SCR{F}[U_0(0,t) u (t)] (t,x/\zeta _2 (t)) + R(t,x)
\end{align*}
with
\begin{align*}
R(t,x) = \frac{e^{i (\zeta _2 '(t)/(2 \zeta _2 (t)) ) x^2}}{(4\pi i b(t))^{n/2}} \int e^{-i x \cdot y /(2b(t))} \left\{
e^{i (\zeta _1(t)/(2 \zeta _2 (t)) ) y^2} -1
\right\}(U_0(0,t) u)(y) dy.
\end{align*}
Here, for any $0 < \alpha <1$,
\begin{align*}
\left|
e^{i (\zeta _1(t)/(2 \zeta _2 (t)) ) y^2} -1 \right| \leq 2 \left|
\sin \left(
\frac{\zeta _1 (t)}{2 \zeta _2(t)} y^2
\right)
\right| \leq C \left| \frac{\zeta _1 (t)}{ \zeta _2(t)}\right|^{\alpha} |y|^{2 \alpha}
\end{align*}
holds. Therefore, we get
\begin{align*}
\left|
e^{i (\zeta _1(t)/(2 \zeta _2 (t)) ) y^2} -1 \right| \leq C \left| \frac{\zeta _1 (t)}{ \zeta _2(t)}\right|^{\alpha} |y|^{2 \alpha}.
\end{align*}
Hence $R(t,x)$ can be estimated as
\begin{align*}
\left\|
R(t)
\right\|_{\infty} &\leq C |\zeta _2 (t)|^{-n/2} \left| \frac{\zeta _1 (t)}{ \zeta _2(t)}\right|^{\alpha} \left\|
|y|^{2 \alpha} U_0(0,t) u(t,y)
\right\| _1 \\ & \leq
C  |\zeta _2 (t)|^{-n/2}  \left| \frac{\zeta _1 (t)}{ \zeta _2(t)}\right|^{\alpha} \left\| U_0(0,t)u(t) \right\|_{0, \gamma}
\end{align*}
for $|t| \geq r_0$ and $\gamma > n/2 + 2 \alpha$.

}

\subsection{MDMDFM decomposition} 
In this subsection, we present the MDMDFM decomposition, which was proposed in \cite{Ko} and modified in \cite{AK} (see \S{7}) and \cite{Ka3} (see, Lemma 2.2. of \cite{Ka3}). This decomposition is used to prove local well-posedness. For the proof of local well-posedness, we employ the Sobolev inequality, that is, 
\begin{align*}
\left\| 
u (t)
\right\|_{\infty} \leq C \left\| u(t) \right\|_{\gamma ,0}. 
\end{align*}     
However it is difficult to obtain an estimate for $\left\| \cdot \right\|_{{H}^{\gamma ,0}}$, which also played a very important role in \cite{HN}. The reason for this difficulty is because if we calculate $pU_0(t,0)$, it satisfies
\begin{align*}
pU_0(t,0) = U_0(t,0) \left(
\frac{1+ \zeta _1' \zeta_2 }{\zeta_1} p + \zeta _1 ' x
\right) = U_0(t,0) \left( \zeta _2 'p + \zeta _1 ' x \right)
\end{align*}
Therefore, the term associated with $x$ appears again. This problem arises because of non-commutativity of $p$ and the propagator. Hence, it difficult to include the norm $\left\| \cdot \right\|_{\gamma ,0}$ in the function space on which the principle of contraction mapping can be applied(see, \eqref{12}). A simple idea for overcoming this difficulty is that 
\begin{align} \nn
\left\| 
u (t)
\right\|_{\infty} &= \left\| e^{-i\zeta _2 '(t) x^2/(2 \zeta _2 (t))}  u(t) \right\|_{\infty} \leq C \left\|e^{-i\zeta _2 '(t) x^2/(2 \zeta _2 (t))}  u(t) \right\|_{\gamma ,0}       \\ & = |\zeta _2(t)|^{-\gamma /2} \left\| |\alpha (t)|^{\gamma /2} u(t) \right\|_2.  \label{13}
\end{align} 
The same function space in \eqref{12} of \S{3} works well. However, $\zeta _2 (t)$ includes some zero points on $ t \in [-r_0,r_0]$ and therefore \eqref{13} fails on these points. In particular, we never remove the condition $\zeta_2 (0) =0$, and therefore it is difficult to prove well-posedness near $t=0$. We believe that this difficulty occurs because the decomposition \eqref{korotya} and \eqref{decomp} exhibit singularities in $\zeta _2 (t) =0$. On the other hand, the decomposition of $U_0(t,0)$ without singularities was also proven in \cite{Ko} and \cite{AK}, and therefore we use such a decomposition to prove local well-posedness; 

We define $a_1(t)$ and $a_2(t)$ as
\begin{align}
&\begin{cases}\label{a1}
a_1(t) = (y_1'(t)y_2(t) - y_1(t)y_2'(t))/(y_1(t)^2 + y_2(t)^2), \\
a_1(0) =1,
\end{cases}\\
&\begin{cases}\label{a2}
a_2(t) = -(y_1'(t)y_1(t) + y_2(t)y_2'(t))/(y_1(t)^2 + y_2(t)^2), \\
a_2(0) =0,
\end{cases}
\end{align}
where $y_1(t)$ and $y_2(t)$ are the solutions to \eqref{2}. We then have the following lemma; 
\begin{Lem}\label{L03} Let $a_1(t)$ and $a_2(t)$ be defined as in \eqref{a1} and \eqref{a2}, respectively. The propagator for $U_0(t, 0)$ then exhibits the following factorization;
\begin{align*}
U_0(t, 0) = \CAL{M} \left( \frac{-1}{a_2(t)} \right)(i)^{n/2} \CAL{D} \left(\frac{1}{\sqrt{a_1(t)}}\right) e^{-i\int_0^t a_1(\tau)d\tau(p^2 + x^2)/2}.
\end{align*}
\end{Lem}  
A merit of this decomposition is that each component of the decomposition exhibits no singular point for any $t \in [-T,T]$, $T \geq r_0$; 
\begin{Lem}\label{L02}
Let $T \geq r_0$ be a large constant. Then for all $t \in [-T, T]$, there exists a constant $ C_{T} >0 $ such that 
\begin{align*}
 |a_1(t)| \leq C_{T}, \quad |a_1(t)|^{-1} \leq C_{T} \quad |a_2(t)| \leq C_{T}.  
\end{align*}
\end{Lem}
The proof can be observed in Lemma 2.2. of \cite{Ka3} and therefore we omit the proof. Using Lemma \ref{L03} and \ref{L02}, we prove the local well-posedness.

\section{Proof of local well-posedness}
In this section, we shall prove the local well-posedeness of the solution to \eqref{eq1} in the $L^{\infty}$ sense. In this section, we consider the assumptions \ref{A1}, \ref{A2}, and \ref{A4}. Let $T \geq r_0$ be a large number and prove the local well-posedness on $t \in [-T,T]$. For simplicity, we only consider the case where $t\in[0,T]$ because the case where $t \in [-T,0]$ can be proven in the same way. $\tilde{C}_{T}$ is a finite positive constant and depends only on $T$. To prove this, we introduce the $L_{T}$ norm and function space $L_{T,M}$ as follows
\begin{align*}
\left\| \hspace{-0.03cm} | \phi \right\| \hspace{-0.095cm}|_{L_{T}} &= \sup_{t\in[0, T]} \| (p^2+x^2+1)^{\gamma /2} \phi (t) \|_2  + \sup_{t\in[0, T]}  \| \phi (t) \|_{\infty} ,
\end{align*}
and for some $M >0$,  
\begin{align*}  
L_{T,M} := 
\Big\{
\phi \in C\left(
[0, T] ; \SCR{S}'
\right);\left\| \hspace{-0.03cm} | \phi \right\| \hspace{-0.095cm}|_{L_{T}} \leq M
\Big\}.
\end{align*}

\begin{Thm}\label{local}
Let $T$ be an enough large so that $T \geq  r_0$ and fixed. Assume $\left\| u_0 \right\|_{\gamma ,0} + \left\| u_0 \right\|_{0, \gamma} = \ep ' \leq \ep$ for some sufficiently small $\ep >0$, where $ \gamma > 0$. Then there exists an $\ep = \ep (T)$, a finite positive constant $C(T)>0$, which depends only on $T$, and a unique solution to \eqref{eq1} such that
\begin{align} \label{ho0}
\left\| \hspace{-0.03cm} | u \right\| \hspace{-0.1cm}|_{L_{T}} \leq C(T) \ep .
\end{align}
Moreover for all $0< \ep ' < \ep$, 
\begin{align} \label{ho1}
\left\| \hspace{-0.03cm} | u \right\| \hspace{-0.1cm}|_{L_{T}} \leq C(T) \ep ' .
\end{align}
\end{Thm}
\Proof{
First, using the commutator calculation, we have
\begin{align*}
\CAL{D}\left(\frac{1}{\sqrt{a_1(t)}}\right)^{-1}x\CAL{D}\left(\frac{1}{\sqrt{a_1(t)}}\right) = \frac{x}{\sqrt{a_1(t)}},\\
\CAL{D}\left(\frac{1}{\sqrt{a_1(t)}}\right)^{-1}p\CAL{D}\left(\frac{1}{\sqrt{a_1(t)}}\right) = \sqrt{a_1(t)}p ,
\end{align*}
and therefore, for $\phi \in \SCR{S}({\bf R}^n)$ and $t, s \in [0, T]$, we obtain 
\begin{align*}
&\| (p^2+x^2 +1)^{\gamma /2} U_0(t, s)\phi \|_2 \\
&= \left\| (p^2+x^2+1)^{\gamma /2} \CAL{M} \left( \frac{-1}{a_2(t)} \right)(i)^{n/2} \CAL{D} \left(\frac{1}{\sqrt{a_1(t)}}\right)  \right. \\ & \left. \qquad \times
e^{-i\int_s^t a_1(\tau)d\tau(p^2 + x^2)/2} \CAL{D}\left(\frac{1}{\sqrt{a_1(s)}}\right)^{-1}(-i)^{n/2}\CAL{M} \left( \frac{1}{a_2(s)} \right)\phi \right\|_2 \\
&= \left\| ((p - a_2(t)x)^2+x^2+1)^{\gamma /2} \CAL{D} \left(\frac{1}{\sqrt{a_1(t)}}\right)  \right. \\ & \left. \qquad \times
e^{-i\int_s^t a_1(\tau)d\tau(p^2 + x^2 )/2} \CAL{D}\left(\frac{1}{\sqrt{a_1(s)}}\right)^{-1}(-i)^{n/2}\CAL{M} \left( \frac{1}{a_2(s)} \right)\phi \right\|_2 \\
&= \left\| \left(\left(\sqrt{a_1(t)}p - \frac{a_2(t)}{\sqrt{a_1(t)}}x\right)^2+\frac{x^2}{a_1(t)}+ 1\right)^{\gamma /2} (p^2+x^2 +1 )^{-\gamma /2} (p^2+x^2 +1 )^{\gamma /2}\right. \\ & \left. \qquad \times
e^{-i\int_s^t a_1(\tau)d\tau(p^2 + x^2)/2} \CAL{D}\left(\frac{1}{\sqrt{a_1(s)}}\right)^{-1}(-i)^{n/2}\CAL{M} \left( \frac{1}{a_2(s)} \right)\phi \right\|_2.
\end{align*}
By Lemma \ref{L01} and Lemma \ref{L02}, the last term of the above equation is smaller than
\begin{align*}
&\tilde{C}_{T} \left\| (p^2+x^2+1)^{\gamma /2} e^{-i\int_s^t a_1(\tau)d\tau (p^2 + x^2)/2} \CAL{D}\left(\frac{1}{\sqrt{a_1(s)}}\right)^{-1}(-i)^{n/2}\CAL{M} \left( \frac{1}{a_2(s)} \right)\phi \right\|_2. 
\\ &= \tilde{C}_{T} \left\| (p^2+x^2+1)^{\gamma /2} \CAL{D}\left(\frac{1}{\sqrt{a_1(s)}}\right)^{-1}(-i)^{n/2}\CAL{M} \left( \frac{1}{a_2(s)} \right)\phi \right\|_2,
\end{align*}
where $p^2+x^2 +1 $ commutes with $e^{-i\int_s^t a_1(\tau)d\tau (p^2 + x^2)/2} $ and $e^{-i\int_s^t a_1(\tau)d\tau (p^2 + x^2)/2} $ is unitary on $L^2({\bf R}^n)$. 
A similar calculation as above yields the following
\begin{align}\label{K2}
&\| (p^2+x^2+1)^{\gamma /2} U_0(t, s)\phi \|_2  \leq \tilde{C}_{T} \left\| (p^2+x^2+1)^{\gamma /2} \phi \right\|_2.
\end{align}
 Now we prove this theorem by employing the contraction mapping principle. Let $\Xi$ as 
 \begin{align*}
(\Xi(u))(t):= U_0(t,0)u_0-i\int_{0}^{t} U_0(t, s)(\nu F_L(u(s))u(s)+\mu F_S(u(s))u(s))ds
\end{align*} 
Then it is enough to show that \\ 
{\bf (I) }. $\Xi$ maps $L_{T,M}$ to $L_{T,M}$. \\ 
{\bf (II)}. $\Xi$ becomes contraction map, i.e., $\Xi$ satisfies 
\begin{align*}
^{\exists}\alpha\in[0, 1)\ {\mathrm s.t.}\ ^{\forall}u,v\in L_{T,M};\quad \left\| \hspace{-0.03cm} | \Xi(u)-\Xi(v) \right\| \hspace{-0.095cm}|_{L_{T}} \leq \alpha \left\| \hspace{-0.03cm} | u-v \right\| \hspace{-0.095cm}|_{L_T}.
\end{align*}
(I). First, we assume that $u \in L_{T,M}$, i.e., for all $t \in [0, T ]$ 
 $$
  \| (p^2+ x^2 +1)^{\gamma /2}  u(t) \|_{2} \leq M, \quad  \|  u(t) \|_{\infty} \leq M,  
 $$ and estimate
 \begin{align*}
  \| (p^2 + x^2 +1)^{\gamma /2}  \Xi  [u] (t) \|_2 .  
 \end{align*}
By the definition of $\Xi$ we estimate
\begin{align*}
& \left\| (p^2 + x^2 +1)^{\gamma /2} \Xi  [u] (t) \right\|_{2} \\ & \leq \| (p^2+x^2+1)^{\gamma /2}U_0(t, 0)u_0 \|_2 + \int_{0}^t\| (p^2+x^2+1)^{\gamma /2} U_0(t, s) G(u(s)) \|_2 ds \\ & \leq
\tilde{C}_{T} \| (p^2+x^2+1)^{\gamma /2}u_0 \|_2 + \tilde{C}_{T} \int_{0}^t\| (p^2+x^2+1)^{\gamma /2} G(u(s)) \|_2 ds, 
\end{align*}
where $G(u(s)) = \nu F_L(u(s)) u(s) + \mu F_S(u(s)) u(s)$.
By Assumption \ref{A4} and Lemma \ref{L01}, we obtain
\begin{align*}  
\| (p^2+x^2+1)^{\gamma /2} F_L(s) u(s) \|_2 & \leq {C} ( \| \J{p}^\gamma  F_L(u(s)) u(s) \|_2 + \| \J{x}^\gamma F_L(u(s)) u(s) \|_2 )\\ & \leq
C ( \tilde{F}_L ( \|u(s)\|_{\infty})  \| \J{p}^\gamma u(s) \|_2 + \tilde{F}_L(  \|u(s)\|_{\infty} ) \| \J{x}^\gamma u(s) \|_2 )\\ & \leq
C \tilde{F}_{L} ( \|u(s)\|_{\infty} ) \| (p^2+x^2+1)^{\gamma /2} u(s) \|_2 \\ 
& \leq C M^{\rho_L} \| (p^2+x^2+1)^{\gamma /2} u(s) \|_2 
,
\end{align*}
and therefore, for $C_M = |\nu| M ^{\rho_L}+ |\mu| M ^{ \rho _S}$
\begin{align} \nn 
& \| (p^2+x^2+1)^{\gamma /2} \Xi_{}[u](t) \|_2  \leq \tilde{C}_{T} \left\| (p^2+x^2+1)^{\gamma /2} u_0 \right\|_{2} \\ & \label{K1}\quad + \tilde{C}_{T} \int_{0}^t \left( 
|\nu| \tilde{F}_L (\left\| u(s) \right\|_{\infty} ) + |\mu | \tilde{F}_S (\left\| u(s) \right\|_{\infty}) \right)  \left\| (p^2+x^2 +1)^{\gamma /2} u(s) \right\|_{2}
ds \\ & \leq 
\tilde{C}_{T} \left\| (p^2+x^2+1)^{\gamma /2} u_0 \right\|_{2} +  C_M \tilde{C}_{T} \int_{0}^t  \left\| (p^2+x^2 +1)^{\gamma /2} u(s) \right\|_{2}
ds \nn \\ & \leq 
\left( \tilde{C}_{T} \ep + C_M  \tilde{C}_{T} T M \right)  \nn
\end{align}
holds. 
Next estimate 
\begin{align*}
  \| \Xi ( u )(t) \|_{\infty} .
\end{align*}
Using Sobolev's inequality, for all $\psi \in \D{(p^2 + x^2 + 1 )^{\gamma}}$, we obtain, 
\begin{align}\label{local2}
\| \psi \|_{\infty}  \leq C \| \J{p}^{\gamma} \psi \|_2  \leq
C \| (p^2+x^2+1)^{\gamma /2} \psi \|_2  . 
\end{align}
Hence, by the same calculations in \eqref{K1} we have 
\begin{align*}
  \left\| \hspace{-0.03cm} | \Xi(u) \right\| \hspace{-0.095cm}|_{L_{T}} \leq 
\left( \tilde{C}_{T} \ep + C_M \tilde{C}_{T} T M \right)  .
\end{align*} 
Consequently, we get 
 \begin{align*}
\left\| \hspace{-0.03cm} | \Xi(u) \right\| \hspace{-0.095cm}|_{L_{T}}  \leq  \tilde{C}_{T}\left(\ep  + C_M  T  M \right)  .
\end{align*}
Here $C_M TM$ is smaller than $C (|\mu| M^{1 + \rho_S} + |\nu| M^{1 + \rho_L} )T$ and hence by putting $M = C(T) \ep $ for some $C(T)>0$, we have 
\begin{align*}
\left\| \hspace{-0.03cm} | \Xi(u)\right\| \hspace{-0.095cm}|_{L_{T}} \leq \tilde{C}_{T} \ep  + C\tilde{C}_{T} \left( 
 C(T)^{1+\rho_L} + C(T)^{1 +\rho_S}
 \right)   \ep  \left( 
|\nu| (\ep )^{\rho_L} + |\mu| (\ep )^{\rho_S}
 \right) T .
\end{align*}
Let $C (T)$ be $C(T) = 4 \tilde{C}_{T}$ and $T >0$ be so that 
\begin{align*}
T(C\tilde{C}_{T} (C(T)^{\rho_L} + C(T)^{\rho_S} )) <  \frac{1}{4 ( 
|\nu| (\ep )^{\rho_L} + |\mu| (\ep )^{\rho_S}
 ) }. 
\end{align*} 
Then for any fixed large $T> r_0 >0$, there exists $\ep = \ep (T)$ such that the above inequality is fulfilled by taking $\ep >0$ sufficiently small compared to $C(T)$ and some constants. For such $C(T)$ and $T$, we notice 
\begin{align*}
\left\| \hspace{-0.03cm} | \Xi(u) \right\| \hspace{-0.095cm}|_{L_{T}} \leq \frac{M}{4} + \frac{M}{4} = \frac{M}{2}.
\end{align*}
Then means $\Xi: L_{T,M} \to L_{T,M}$. 
\\ ~~ \\ 
(II). By using \eqref{est} and the same argument as in Proposition \ref{prop2.2} and Proposition \ref{P1}, we get 
\begin{align*}
& \left\| (p^2 + x^2 +1) \left(  
G(u(s)) - G(v(s))  \right) 
\right\|_{2} \\ & \leq C \left( 
|\nu| \left\| 
F_L(u(s))
\right\|_{\infty} + |\nu| \left\| 
F_L(v(s))
\right\|_{\infty} + |\mu| \left\| 
F_S(u(s))
\right\|_{\infty} + |\mu| \left\| 
F_S(u(s))
\right\|_{\infty}  
\right) \\ & \qquad \times  \left\| 
(p^2 + x^2 +1) (u(s)-v (s))
\right\|_{2}.
\end{align*}
 By using this inequality and the same argument as in the proof for (I), we have 
 \begin{align*}
 \left\| 
 (p^2 + x^2 +1) \left( \Xi (u) (t) - \Xi (v) (t) \right) 
 \right\|_{2} \leq C_M \tilde{C}_T T  \sup_{t \in [0,T]}\left\| (p^2 + x^2+1 ) (u(t) -v(t))\right\| _2
 \end{align*}
 and 
 \begin{align*}
 \left\| 
 \Xi (u) (t) - \Xi (v) (t)
 \right\|_{\infty} \leq C_M \tilde{C}_T T  \sup_{t \in [0,T]}\left\| (p^2 + x^2+1 ) (u(t) -v(t))\right\| _2. 
 \end{align*} 
Hence we get 
\begin{align*}
\left\| \hspace{-0.03cm} | \Xi(u)-\Xi(v) \right\| \hspace{-0.095cm}|_{L_{T}}\leq C (|\nu | M^{\rho_L} + |\mu | M^{\rho_S} ) \tilde{C}_T  T \left\| \hspace{-0.03cm} | u-v \right\| \hspace{-0.095cm}|_{L_{T}} . 
\end{align*}
By taking $M = C(T) \ep ' $ and supposing
\begin{align*}
T(C \tilde{C}_T (C(T)^{\rho_L} + C(T)^{\rho_S} )) \leq \frac{1}{2(|\nu| (\ep )^{\rho_L} + |\mu| (\ep )^{\rho_S})}, 
\end{align*}
we get 
\begin{align*}
\left\| \hspace{-0.03cm} | \Xi(u)-\Xi(v) \right\| \hspace{-0.095cm}|_{L_{T}} \leq \frac{1}{2}
\left\| \hspace{-0.03cm} | u-v \right\| \hspace{-0.095cm}|_{L_{T}} 
. 
\end{align*}
By using the contraction mapping theorem (see, e.g., Cazenave \cite{Caz}), we have \eqref{ho0}. Moreover by taking $M= C(T) \ep '$ and by using $\ep ^{-1} < (\ep ')^{-1}$ we also have \eqref{ho1}. These complete the proof.
}

\section{Proof of Theorem \ref{thm} and \ref{T2}.}
In this section, we prove Theorem \ref{thm} with $|t| \geq r_0$. Then using Theorem \ref{local}, we can derive Theorem \ref{thm}. Theorem \ref{T2} can be obtained as a sub-consequence of Theorem \ref{thm}. Similarly in \S{3}, we only consider the case where $t \geq r_0$. In this section, we always consider assumptions \ref{A1}, \ref{A2}, and \ref{A4} and if necessary, we assume assumption \ref{A3} additionally. Let $T> r_0$ be the same one given in \S{3}. We define the function space $X_T$ as follows
\begin{align}  
X_T := \Big\{
\phi \in C\left(
[r_0, T] ; \SCR{S}'
\right); \left\| \hspace{-0.03cm} | \phi \right\| \hspace{-0.095cm}|_{X_T} &= \label{12}
\CB{\sup_{t\in[r_0,T]} \left( (1+ t)^{C_1(\ep ',\nu) }  +  e^{C_2(\ep', \mu)} \right)^{-1} \| U_0(0,t)\phi (t) \|_{0, \gamma}}  \\ 
& \qquad + \sup_{t\in[r_0,T]} (1+|\zeta _2 (t)|)^{n/2} \| \phi (t) \|_{\infty} < \infty
\Big\}, \nn 
\end{align}
where \CB{$C_1(\ep ' , \nu), C_2 (\ep ', \mu) >0$ are sufficiently small constants that are later included.} 

We set $u(t,x)$ is the solution to \eqref{eq1}. Then according to Theorem \ref{local}, we have 
\begin{align*}
\left\| u(r_0) \right\|_{\gamma,0} \leq C(r_0) \ep ' , \qquad \left\| u(r_0) \right\|_{0, \gamma } \leq C(r_0) \ep ' 
\end{align*}
under the assumption $\left\| u_0 \right\|_{\gamma ,0} + \left\| u_0 \right\|_{0, \gamma} = \ep ' \leq \ep$. Because $r_0$ is a given constant, we assume that $\ep ' >0$ is sufficiently small compared to $\tilde{C}_{r_0}$, i.e., $ 0< C(r_0) \ep ' \ll 1$. From this, we rewrite $C(r_0)$ as $C$ and in the following, we assume 
\begin{align*}
\left\| u(r_0) \right\|_{\gamma,0} \leq C \ep ' , \qquad \left\| u(r_0) \right\|_{0, \gamma } \leq C \ep '. 
\end{align*}
Then by the same calculations in \S{3}, we can also obtain 
\begin{align}\label{14}
\left\| U_0(0,r_0) u(r_0) \right\|_{\gamma,0} \leq C \ep ' , \qquad \left\| U_0(0,r_0) u(r_0) \right\|_{0, \gamma } \leq C \ep '.
\end{align} 

Furthermore we set the following lemma to prove the Theorem \ref{thm}.
\begin{Lem}\label{lem4.1}
Assume $\left\| u_0 \right\|_{\gamma ,0} + \left\| u_0 \right\|_{0, \gamma} = \ep ' \leq \ep$ for some sufficiently small $\ep >0$, where $ \gamma  >n/2$. Then there exists a finite interval $[r_0,T]$ and a unique solution to \eqref{eq1} such that
\begin{align*}
\left\| \hspace{-0.03cm} | u \right\| \hspace{-0.095cm}|_{X_T} \leq T_{\ep'}, 
\end{align*}
where a constant $T_{\ep '} >0$ satisfies that for given $T$, $T_{ \ep '} \to 0$ as $\ep ' \to 0$.
\end{Lem} 
\Proof{
Since $\left\| \hspace{-0.03cm} | u \right\| \hspace{-0.095cm}|_{L_{T}} \leq C_T \ep ' $ holds, we have 
\begin{align*}
&  \left( (1+ t)^{C_1(\ep ',\nu) }  +  e^{C_2(\ep', \mu)} \right)^{-1} \| U_0(0,t) u (t) \|_{0, \gamma} \\ & \leq C \left\| (p^2 + x^2 +1) U_0(0,t) u (t) \right\|_2 \leq C_T  \left\| (p^2 + x^2 +1) u (t) \right\|_2 \leq C_T \ep '
\end{align*}
and 
\begin{align*}
(1 + |\zeta _2 (t)|)^{n/2} \left\| u(t) \right\|_{\infty} \leq C_T \left\| (p^2 + x^2 +1) u(t) \right\|_{2} \leq C_T \ep '.
\end{align*} 
}

\begin{Thm}\label{Tz1}
Let $u$ be the local solutions to \eqref{eq1} stated in Lemma \ref{lem4.1}. Then for any $t \in [-T, -r_0] \cup [r_0 , T]$, 
\begin{align*}
(1+ |\zeta _2 (t)|)^{n/2} \| u(t) \|_{\infty} \leq C \ep' , 
\end{align*}
where a constant $C>0$ does not depend on $T$.
\end{Thm}

\Proof{ First, for $0 \leq s \leq r_0$, we use the obtained result in Theorem \ref{local}
 \begin{align*}
(1+s) \tilde{f}(\| u(s) \| _{\infty}) \leq C (\ep ')^{\rho_S} \mbox{ or } C (\ep ')^{\rho_L} 
 \end{align*} 
 and for $s \geq r_0$, obtained result in Lemma \ref{lem4.1}
 $$
 \|  u(s) \|_{\infty} \leq T_{\ep '} (1 + |\zeta _2 (s)| )^{-n/2} \leq T_{\ep '} |\zeta _2 (s)|^{-n/2}. 
 $$ Then we estimate
 \begin{align*} 
\sup_{t \in [r_0,T]}
\left( 
(1+ t)^{C_1 (\ep ', \nu) } + e^{C_2 (\ep ', \mu)}
\right)^{-1} 
 \left\| U_0(0,t) u(t)\right\|_{0, \gamma}
\leq C \ep' , 
 \end{align*} 
 where $C>0$ is independent of $T$.
 By using Duhamel's formula, we have 
\begin{align*}
\left\| U_0(0,t)u(t)\right\|_{0, \gamma} & \leq
\left\|  u_0 \right\|_{0, \gamma} + \int_{0}^t \left\| \J{x}^{\gamma} U_0(0,s) G(u(s)) \right\|_2 ds.
\end{align*}
According to proposition \ref{prop2.2}, we obtain
\CB{\begin{align*}
& \left\| U_0(0,t)u(t)\right\|_{0, \gamma}\\  & \leq
\left\| u_0 \right\|_{0, \gamma}  +
\int_{0}^t \left( |\nu| \tilde{F}_L ( \left\| u (s) \right\|_{\infty} ) + |\mu| \tilde{F}_S ( \left\| u (s) \right\|_{\infty} ) \right) \left\| U_0(0,s) u(s) \right\|_{0, \gamma}ds 
\\ & \leq \left\| u_0 \right\|_{0, \gamma} + C 
\int_{0}^t \left( |\nu|(T_{\ep '})^{\rho_L} (1+s)^{-1} + |\mu |(T_{\ep '})^{\rho_S} (1+s)^{-1-\delta_0} \right) \left\| U_0(0,s) u(s) \right\|_{0, \gamma} ds.
\end{align*}}
\CG{Define $ C|\nu| (T_{\ep '})^{\rho_L} =: C_1 (\ep ',\nu)$ and $ C|\mu| (T_{\ep '})^{\rho_S} =: C_2 (\ep ',\mu)$}. Then using the Gronwall inequality, we obtain
\begin{align}\label{est4}
\left\| U_0(0,t)u(t)\right\|_{0, \gamma} \leq C \CB{\left(  (1+t)^{C_1 (\ep ',\nu)} + e^{C_2(\ep ', \mu)} \right) \left\| u_0 \right\|_{0,\gamma}}.
\end{align}

Next we estimate $\|u (t) \| _{\infty}$. From lemma \ref{lem2.6}, it holds that for $t \geq r_0$
\begin{align}\label{est6}
\left\|
u(t)
\right\|_{\infty} & \leq
C \CB{\ep '}(1+ |\zeta _2 (t)|)^{-n/2} \left| \frac{\zeta _1 (t)}{\zeta _2 (t)} \right|^{\alpha} \CB{ \left(  (1+ t)^ { C_1(\ep ',\nu)}   + e^{C_2(\ep',\mu)} \right)} \left\|  u_0 \right\|_{ 0, \gamma} \notag \\ & \quad + (1+ |\zeta _2 (t)| )^{-n/2} \left\| \SCR{F} U_0(0,t) u(t) \right\|_{\infty}.
\end{align}
and therefore we shall estimate the term $(1+|\zeta _2 (t)|)^{-n /2} \left\| \SCR{F} U_0(0,t) u(t) \right\|_{\infty}$ in the following manner; 

 Through simple calculations, it holds that
\begin{align}\label{eq2}
i \partial _t \left(
U_0(0,t) u(t)
\right) = U_0(0,t) G(u(t)) = U_0(0,t) \left(
\nu {F}_L(u(t)) u{(t)} + \mu {F}_S (u(t)) u{(t)}
\right) .
\end{align}
The term $ U_0(0,t) {F}_L (u(t)) u (t) $ is estimated as
\begin{align}\label{eq3}
& U_0(0,t) {F}_L(u(t)) u(t) \notag \\ &=
\CAL{M} \left(
 -\frac{\zeta _2(t)}{\zeta _1 (t) } \right) \SCR{F}^{-1} \CAL{D}(\zeta _2 (t))^{-1} \CAL{M} \left( - \frac{\zeta _2 (t)}{ \zeta _2 '(t)} \right)
 {F}_L(u(t)) u(t) \notag \\ &=
\CAL{M} \left(
 -\frac{\zeta _2(t)}{\zeta _1 (t) } \right) \SCR{F}^{-1} \CAL{D}(\zeta _2 (t))^{-1} {F}_L \left( \CAL{M} \left( -\frac{\zeta _2 (t)}{ \zeta _2 '(t)} \right) u(t) \right) \CAL{M} \left( -\frac{\zeta _2 (t)}{ \zeta _2 '(t)} \right)
u(t) \notag \\ &=
 \CAL{M} \left(
- \frac{\zeta _2(t)}{\zeta _1 (t) } \right) \SCR{F}^{-1} F_L \left( |\zeta _2(t)|^{-n/2}
\CAL{D}(\zeta _2 (t))^{-1} \CAL{M} \left(- \frac{\zeta _2 (t)}{ \zeta _2 '(t)} \right) u(t)
\right) \\& \qquad \times  \CAL{D}(\zeta _2 (t))^{-1} \CAL{M} \left( -\frac{\zeta _2 (t)}{ \zeta _2 '(t)} \right) u(t) . \nn
\end{align}
For simplicity, we denote
\begin{align*}
\CAL{M}_1 = \CAL{M} \left(
 -\frac{\zeta _2(t)}{\zeta _1 (t) } \right) , \quad \CAL{D} = \CAL{D}(\zeta _2 (t)), \quad \CAL{M}_2 = \CAL{M} \left( -\frac{\zeta _2 (t)}{ \zeta _2 '(t)} \right) .
\end{align*}
According to $\SCR{F} \CAL{M}_1^{-1} U_0(0,t) = \CAL{D}^{-1} \CAL{M}_2$, the above equation is equivalent to
\begin{align*}
\CAL{M}_1 \SCR{F}^{-1} F_L\left( |\zeta _2(t)|^{-n/2}
\SCR{F} \CAL{M}_1^{-1} U_0(0,t) u{(t)}
\right) \SCR{F} \CAL{M}_1^{-1} U_0(0,t) u (t).
\end{align*}
Here we denote $v(t) := U_0(0,t) u(t)$. The above term is then equivalent to
\begin{align} \nn 
& \Big\{
\left(
\CAL{M}_1 -1
\right) \SCR{F}^{-1} F_L \left(|\zeta _2(t)|^{-n/2} \widehat{\CAL{M}_1^{-1} v} (t) \right) \widehat{\CAL{M}_1^{-1} v} (t) \\ & \qquad + \SCR{F}^{-1}
\left(
F_L\left( |\zeta _2(t)|^{-n/2}\widehat{\CAL{M}_1^{-1} v} (t) \right) \widehat{\CAL{M}_1^{-1} v} (t) - F_L \left(
|\zeta _2(t)|^{-n/2} \hat{v}(t)
\right) \hat{v}(t) \label{X1}
\right)
\Big\} \\ & \qquad \qquad + \SCR{F}^{-1} F_L\left( |\zeta _2(t)|^{-n/2} \hat{v}(t) \right) \hat{v}(t). \nn 
\end{align}
By performing a Fourier transform on both sides of \eqref{eq2}, we obtain
\begin{align*}
i\partial _t \hat{v}(t) = \nu \SCR{F} U_0(0,t) F_L(u(t)) u(t) + \mu \SCR{F} U_0(0,t) F_S( u(t))u(t).
\end{align*}
By using
\begin{align*}
U_0(0,t) F_L(u(t)) u(t) = \eqref{eq3} = \eqref{X1},
\end{align*}
we obtain
\begin{align*}
i \partial _t \hat{v} (t) &= \nu F_L \left(|\zeta _2(t)|^{-n/2} \hat{v}(t) \right) \hat{v}(t) \\ & \quad +
\nu  \left( I_1(t) + I_2 (t) \right) \\ & \qquad + \mu \hat{Q} (t),
\end{align*}
where
\begin{align*}
I_1 (t) &:= \SCR{F} \left( \CAL{M}_1 -1 \right) \SCR{F}^{-1} F_L \left(|\zeta _2(t)|^{-n/2} \widehat{\CAL{M}_1^{-1}v} (t) \right) \widehat{\CAL{M}_1^{-1} v} (t), \\
I_2(t) &:= F_L \left(|\zeta _2(t)|^{-n/2} \widehat{\CAL{M}_1^{-1}v} (t) \right) \widehat{\CAL{M}_1^{-1} v} (t) -
F_L\left( |\zeta _2(t)|^{-n/2} \hat{v}(t) \right) \hat{v}(t)
\end{align*}
and
\begin{align*}
{Q}(t) := U_0(0,t) F_S\left( {u}(t) \right) {u}(t).
\end{align*}
Let us define
\begin{align*}
\hat{w}(t) := B(t) \hat{v} (t), \quad
B(t) := \exp \left( {i \nu \int_{r_0}^t {\left(
F_L\left( |\zeta _2(\tau)|^{-n/2} \hat{v}(\tau) \right) 
\right) }d\tau } \right).
\end{align*}
Then $\hat{w}$ satisfies
\begin{align*}
i \partial _t \hat{w}(t) = \nu B(t)  \left( I_1(t) + I_2(t) \right) + \mu B(t) \hat{Q}(t).
\end{align*}
Integrating both sides in $t$ from $r_0$ to t,
\begin{align*}
\hat{w}(t) - \hat{w}(r_0) = -i \int_{r_0}^t B (\tau) \left\{ \nu  \left(
I_1(\tau) + I_2 (\tau)
\right) + \mu \hat{Q}(\tau) \right\} d \tau,
\end{align*}
and therefore
\begin{align}\label{eq4}
& \left\| \SCR{F} U_0(0,t) u(t) \right\|_{\infty} \\  &= \left\| \hat{w}(t) \right\|_{\infty} \notag \\ & \leq C \ep ' + C\int_{r_0}^t \left\{ |\nu | \left( \left\| I_1 (\tau) \right\|_{\infty} + \left\| I_2(\tau) \right\| _{\infty} \right) 
+ | \mu | \left\| \hat{Q}(\tau)\right\|_{\infty} \right\} d \tau \nn
\end{align}
holds, when $\gamma '>n/2$,
\begin{align*}
\left\| 
\hat{w}(r_0)
\right\|_{\infty} = \left\| \SCR{F} U_0(0,r_0) u(r_0) \right\|_{\infty} \leq C \left\|  U_0(0,r_0) u(r_0) \right\|_{1} \leq C \left\| U_0(0,r_0) u(r_0) \right\|_{0, \gamma'} \leq C \ep'
\end{align*} 
We now estimate each term in integration;
\begin{align*}
\left\| 
I_1(t)
\right\|_{\infty} &= \left\| \SCR{F} \left( \CAL{M}_1 -1 \right) \SCR{F}^{-1} F_L \left( |\zeta _2 (t)|^{-n/2}\widehat{\CAL{M}_1^{-1}v} \right) \widehat{\CAL{M}_1^{-1} v} \right\|_{\infty} \\ & \leq
C \left\| \left( \CAL{M}_1 -1 \right) \SCR{F}^{-1} F_L\left(|\zeta _2 (t)|^{-n/2} \widehat{\CAL{M}_1^{-1} v} \right) \widehat{\CAL{M}_1^{-1} v} \right\|_{1} \\ & \leq
C \left| \frac{\zeta _1 (t)}{\zeta _2 (t)} \right|^{\alpha} \left\| |x|^{2 \alpha} \SCR{F}^{-1} F_L\left( |\zeta _2 (t)|^{-n/2} \widehat{\CAL{M}_1^{-1} v} \right) \widehat{\CAL{M}_1^{-1} v} \right\|_{1},
\end{align*}
where $0< \alpha <1$. Here by using Schwarz's inequality and Lemma \ref{lem2.6}, for $\gamma >n/2 + 2 \alpha$ and $\gamma ' > n/2$, the above inequality is smaller than
\begin{align*}
& C \left| \frac{\zeta _1 (t)}{\zeta _2 (t)} \right|^{\alpha} \left\| \SCR{F}^{-1} F_L \left(|\zeta _2 (t)|^{-n/2} \widehat{\CAL{M}_1^{-1}v} \right) \widehat{\CAL{M}_1^{-1} v} \right\|_{0,\gamma} \\ \leq & C
\left| \frac{\zeta _1 (t)}{\zeta _2 (t)} \right|^{\alpha} |\zeta _2 (t)|^{n/2}\left\| F_L \left( |\zeta _2 (t)|^{-n/2} \widehat{\CAL{M}_1^{-1}v} \right) |\zeta _2 (t)|^{-n/2} \widehat{\CAL{M}_1^{-1} v} \right\|_{\gamma ,0} \\ \leq & C
\left| \frac{\zeta _1 (t)}{\zeta _2 (t)} \right|^{\alpha} |\zeta _2 (t)|^{n/2}
\tilde{F}_L\left(  |\zeta _2 (t)|^{-n/2} \left\|  \widehat{\CAL{M}_1^{-1} v} \right\|_{\infty} \right)  |\zeta _2 (t)|^{-n/2} \left\| \widehat{\CAL{M}_1^{-1} v} \right\|_{\gamma, 0} \\ 
\leq & C \left| \frac{\zeta _1 (t)}{\zeta _2 (t)} \right|^{\alpha} |\zeta _2 (t)|^{-n\rho_L/2} 
\left\| {\CAL{M}_1^{-1} v} \right\|_{1}^{\rho_L}  \left\| {\CAL{M}_1^{-1} v} \right\|_{ 0, \gamma} \\
\leq & C \left| \frac{\zeta _1 (t)}{\zeta _2 (t)} \right|^{\alpha}
 |\zeta _2 (t)|^{-n\rho_L/2} 
\left\|  v \right\|_{0, \gamma '}^{\rho_L}  \left\|  v \right\|_{ 0, \gamma}  ,
\end{align*}
where we use \eqref{est4}, and above inequality gives
\begin{align}\label{I1}
\left\|
I_1(t)
\right\|_{\infty} \leq C \left| \frac{\zeta _1 (t)}{\zeta _2 (t)} \right|^{\alpha} |\zeta _2 (t)|^{-n\rho_L/2} 
\left\| {v} \right\|_{0, \gamma '}^{\rho_L}  \left\| { v} \right\|_{ 0, \gamma} 
.
\end{align}
On the other hand,
\begin{align}\label{I2}
\left\|
I_2(t)
\right\|_{\infty} &= \left\| F_L \left(|\zeta _2 (t)|^{-n/2} \widehat{\CAL{M}_1^{-1} v} \right)\widehat{\CAL{M}_1^{-1} v} - F_L \left( |\zeta _2 (t)|^{-n/2} \hat{v} \right) \hat{v} \right\|_{\infty} \notag \\ & \leq C \left(
\tilde{F}_L \left( |\zeta _2 (t)|^{-n/2} \left\| \widehat{\CAL{M}_1^{-1} v} \right\|_{\infty} \right) + \tilde{F}_L \left( |\zeta _2 (t)|^{-n/2}\left\| \hat{v} \right\|_{\infty} \right) 
\right) \left\| \widehat{\CAL{M}_1^{-1} v } -\hat{v} \right\|_{\infty} \notag \\ 
& \leq
C |\zeta _2 (t)|^{-n\rho_L/2} \left\| v \right\|_{1}^{\rho_L} \left\| \left( \CAL{M}_1^{-1} -1\right) v \right\|_1 \notag \\
 & \leq
C \left| \frac{\zeta _1 (t)}{\zeta _2 (t)} \right|^{\alpha}
|\zeta _2 (t)|^{-n\rho_L/2} \left\| v \right\|_{0, \gamma '}^{\rho_L} 
 \left\| v\right\|_{0, \gamma}
\end{align}
and
\begin{align}\label{Q}
\left\|
\hat{Q}(t)
\right\|_{\infty} &= \left\| \SCR{F} U_0(0,t) F_S(u) u \right\|_{\infty} \notag \\ & \leq C
\left\| \SCR{F} \CAL{M}_1 \SCR{F}^{-1} \CAL{D}^{-1} \CAL{M}_2 F_S(u) u \right\|_{\infty}
\notag \\ 
& \leq C \left\| \SCR{F}^{-1} F_S \left( |\zeta _2 (t)|^{-n/2} \widehat{\CAL{M}_1^{-1}v} \right) \widehat{\CAL{M}_1^{-1}v} \right\|_1\notag \\ 
& \leq C |\zeta _2 (t)|^{-n\rho_S/2} \left\| \widehat{\CAL{M}_1^{-1} v}\right\|_{\infty}^{\rho_S} \left\| \widehat{\CAL{M}_1^{-1} v} \right\|_{\gamma ',0} \notag \\ 
& \leq C
 |\zeta _2 (t)|^{-n\rho_S/2} \left\|  v\right\|_{0,\gamma '}^{\rho_S} \left\| v \right\|_{0, \gamma '} 
\end{align}
Using \eqref{eq4} -- \eqref{Q} \CG{and together with
\begin{align*}
 \left\| v(t) \right\|_{0,\gamma} \leq C \ep' \left( 
  (1+t)^{C_1 (\ep ',\nu)} +  e^{C_2(\ep ', \mu)}  \right) \leq C \ep' (1+t)^{C_1 (\ep ',\nu)}
\end{align*} 
for $t \geq r_0$,} we obtain   
\begin{align}
&  \nn \left\|
\SCR{F}U_0(0,t) u(t)
\right\|_{\infty} \\ 
& \nn \leq C \ep ' + \int_{r_0}^t \Bigg\{ | \nu |
\left| \frac{\zeta _1 (\tau)}{ \zeta _2 (\tau) } \right|^{\alpha} |\zeta _2(\tau)|^{-n \rho_L/2}  \left\|  v \right\|_{0, \gamma '}^{{\rho}_L} \left\| v \right\|_{0, \gamma} + | \mu | |\zeta _2 (\tau)|^{-n \rho_S/2} \left\|  v \right\|_{0,\gamma '}^{{\rho}_S} \left\|   v \right\|_{0,\gamma '}
\Bigg\} d \tau \\
 & \nn  \leq
C \ep ' + \CG{ C \ep ' \int_{r_0}^t \Bigg\{
(\ep ')^{\rho_L} | \nu | 
 \left| \frac{\zeta _1 (\tau)}{ \zeta _2 (\tau) } \right|^{\alpha}|\zeta _2 (\tau)|^{-n\rho_L/2} |\tau|^{C_1(\ep ', \nu ) (1+ {\rho}_L) }} \\ 
 & \qquad \nn
 ~~~~~~~~~~~~~~~~~~~~~~~~~~~~~~~~~~~~~~~~~
 \CG{+ (\ep ')^{\rho_S} |\mu|  |\zeta _2(\tau)|^{-n\rho_S/2} |\tau|^{C_1(\ep ', \mu) ( 1 + {\rho}_S) }
\Bigg\} d \tau}
\\ 
& \nn  \leq
C \ep '  + C \ep ' \CG{((\ep ')^{\rho_L} + (\ep ')^{\rho_S})} 
\\ 
& \label{aal} \qquad ~~~~~~~~ \times\int_{r_0}^t \Bigg\{ |\nu |
\left| \frac{\zeta _1 (\tau)}{ \zeta _2 (\tau) } \right|^{\alpha} |\tau|^{-1 + C_1(\ep ',\nu) (1+ {\rho}_L )}  + | \mu | |\tau|^{-1 - \delta _1 + C_2(\ep ',\mu) (1 + {\rho}_S)}
\Bigg\} d \tau.
\end{align}
Here we note $C_1(\ep ', 0) =0$, $C_1 (\ep', \nu) $ and $ C_2 (\ep ', \mu) \to 0$ as $\ep' \to 0$, respectively, and then based on assumption \ref{A2} with $\nu =0$ and $\delta_1 > C_2(\ep',\nu)(1+ {\rho}_S ) $, we obtain the above term, which is smaller than $C \ep '$, and based on assumption \ref{A3} with $\nu \neq0$, $\delta_1 > C_2(\ep',\mu)(1+ {\rho}_S ) $ and $\alpha \delta_0 > C_1(\ep',\nu)( 1+ {\rho}_L) $, we obtain the above inequality, which is smaller than 
\begin{align*}
C \ep' + C\ep' \CG{((\ep ')^{\rho_L} + (\ep ')^{\rho_S})}  \int_{r_0}^t ( |\tau|^{-1-\delta _0 \alpha + C_1(\ep ',\nu) (1+ {\rho}_L ) } +  |\tau|^{-1 - \delta _1 + C_2(\ep ',\mu) (1 + {\rho}_S)}) d \tau 
\end{align*}
and therefore, we finally obtain
\begin{align}\label{est7}
\left\|
\SCR{F}U_0(0,t) u(t)
\right\|_{\infty} \leq C \ep '.
\end{align}
Combining \eqref{est4}, \eqref{est6}, and \eqref{est7}, we finally derive Theorem \ref{Tz1}.}

\subsection{Proof of Theorem \ref{thm} and \ref{T2}} We now prove Theorem \ref{thm}. From Theorem \ref{local} and Theorem \ref{Tz1}, we have 
\begin{align*}
\left\| \hspace{-0.03cm} | u \right\| \hspace{-0.095cm}|_{X_T} \leq C \left( 
\| u_0 \|_{\gamma ,0} + \| u_0 \|_{0, \gamma}
\right)  = C \ep '
\end{align*}
for all $t \in [-T, T]$. Because the constant $C$ does not depend on $T$, we apply the continuation argument and obtain Theorem \ref{thm}. \CG{Moreover, Theorem \ref{T2} can be proven by imitating the proof of Theorem 1.2. in \cite{HN}.} 
By Lemma \ref{lem4.1} and the same calculation in \eqref{K2} and \eqref{K1}, we get 
\begin{align*}
& \left\| 
(p^2 + x^2 +1) ^{\gamma /2} u(t)
\right\|_2  \\ & \leq q(t)  \left\| 
(p^2 + x^2 +1) ^{\gamma /2} u_0
\right\|_2 + q(t) C(\ep ') \int_0^t  \left\| 
(p^2 + x^2 +1) ^{\gamma /2} u(s)
\right\|_2  ds,  
\end{align*}
where $q(t)$ satisfies for $t \geq r_0$ $|q(t) | \leq |\zeta _2 (t)|^{4 \gamma}$. Since $\zeta _2 (t)$ is continuous we have $u(t) \in C({\bf R} ; H^{\gamma ,0} \cap H^{0, \gamma})$ by using Gronwall inequality.

\end{document}